\documentclass{appolb}
\usepackage{amsmath,amssymb,amsfonts}
\usepackage{bbm}
\usepackage{graphicx}

\begin{document}
\title{3D gravity, point particles and deformed symmetries%
\thanks{Presented at {\it The 8th Conference of the Polish Society on Relativity}, Warszawa, Poland, September 19-23, 2022.}%
}
\author{Tomasz Trze\'{s}niewski\footnote{tomasz.trzesniewski@uwr.edu.pl}
\address{Institute of Theoretical Physics, University of Wroc\l{}aw, pl.\ M.\ Borna 9, \\50-204 Wroc\l{}aw, Poland}
\\[3mm]
}


\maketitle
\begin{abstract}
It is well known that gravity in 2+1 dimensions can be recast as Chern-Simons theory, with the gauge group given by the local isometry group, depending on the metric signature and the cosmological constant. Point particles are added into spacetime as (spinning) conical defects. Then, in principle, one may integrate out the gravitational degrees of freedom to obtain the effective particle action; the most interesting consequence is that the momentum space of a particle turns out to be curved. This is still not completely understood in the case of non-zero cosmological constant. 
\end{abstract}
  
\section{Prelude}
The counting of degrees of freedom shows that 3D gravity is a topological theory, with no action at a distance or wave solutions. If no other fields are included, dynamics can only be introduced into it by a nontrivial (spatial) topology or conical defects, interpreted as point particles \cite{Staruszkiewicz:1963ge}. Consequently, depending on the cosmological constant, spacetime is locally isometric to 3D Minkowski or (anti-)de Sitter space. Non-charged black hole solutions exist only in the adS case \cite{Banados:1992te}, as an extension of the family of particle solutions, and they have the topology of a handle. In this short paper, we will restrict to particles living on a (closed) spatial surface of genus 0.

\section{Local isometry groups and spinning conical defects}
Instead of the metric $g_{\alpha\beta}$, gravity can be described in terms of the vielbein $e_\alpha^{\ \mu}$ and spin connection $\omega_\alpha^{\ \mu\nu}$, defined by the formulae
\begin{align}\label{eq:10.01}
e_\alpha^{\ \mu} e_\beta^{\ \nu} \eta_{\mu\nu} = g_{\alpha\beta}\,, \qquad 
\omega_\alpha^{\ \mu\nu} = e_\beta^{\ \mu} \partial_\alpha e^{\beta\nu} + e_\beta^{\ \mu} \Gamma_{\ \alpha\gamma}^\beta e^{\gamma\nu}\,.
\end{align}
In (2+1)D, they neatly combine into a gauge field with values in the local isometry algebra $\mathfrak{g}$ (3D Poincar\'{e} or (anti-)de Sitter, depending on the cosmological constant $\Lambda$), which is the Cartan connection
\begin{align}\label{eq:10.02}
A = -\tfrac{1}{2} \epsilon^\mu_{\ \nu\sigma} \omega_\alpha^{\ \nu\sigma} J_\mu dx^\alpha + e_\alpha^{\ \mu} P_\mu dx^\alpha
\end{align}
and $J_\mu$, $P_\mu$ are generators of $\mathfrak{g}$. As a result \cite{Witten:1988dm}, Einstein-Hilbert action can be rewritten as the action of Chern-Simons theory
\begin{align}\label{eq:10.03}
S = \frac{1}{16\pi G} \int \left(\left<dA \wedge A\right> + \frac{1}{3} \left<A \wedge [A,A]\right>\right)
\end{align}
if the inner product on $\mathfrak{g}$ is given by ($\eta$ denotes Minkowski metric)
\begin{align}\label{eq:10.04}
\left<J_\mu, P_\nu\right> = \eta_{\mu\nu}\,, \qquad 
\left<J_\mu, J_\nu\right> = \left<P_\mu, P_\nu\right> = 0\,.
\end{align}
More generally, the inner product can be a linear combination of (\ref{eq:10.04}) and
\begin{align}\label{eq:10.09}
\left<J_\mu, P_\nu\right>_{\rm alt} = 0\,, \qquad 
\left<J_\mu, J_\nu\right>_{\rm alt} = -\Lambda^{-1} \left<P_\mu, P_\nu\right>_{\rm alt} = \eta_{\mu\nu}
\end{align}
but the theory with the latter turns out to be not the ordinary gravity \cite{Meusburger:2009gy}. 

\subsection{Coupling a particle to Chern-Simons action}
A neighbourhood of a massive point-particle in (2+1)D is described \cite{Deser:1984tc,Deser:1984tr} by the vacuum spacetime-interval
\begin{align}\label{eq:10.05}
ds^2 = (1 - \Lambda r^2)\, dt^2 - (1 - \Lambda r^2)^{-1} dr^2 - r^2 d\tilde\phi^2\,,
\end{align}
if the polar angle is rescaled to $\tilde\phi := (1 - 4G m)\, \phi$ (this is what we call a conical defect). The particle's spin $\neq 0$ induces a similar jump in the time coordinate, hence the geometry becomes that of a helical cone. 

Let spacetime have the topology $\mathbbm{R} \times {\cal S}$. The model of a single particle on a closed surface ${\cal S}$ is not well defined globally for any value of $\Lambda$ but may be used as a step in solving the multiparticle case (as well as models with open ${\cal S}$), see subsec.~\ref{seq:3.2}. Then, the field $A = A_t dt + A_S$ and the action of gravity coupled to a particle (at $\vec{x}_*$) is \cite{Sousa:1990os,Meusburger:2009gy}
\begin{align}\label{eq:10.06}
S = \int\! dt\ L\,, \quad &L = \frac{1}{16\pi G} \int_{\cal S} \left<\dot A_{\cal S} \wedge A_{\cal S}\right> - \left<c_0 h^{-1} \dot h\right> + \nonumber\\ 
&\int_{\cal S} \left<A_t \left(\frac{1}{8\pi G} F_{\cal S} - h c_0 h^{-1} \delta^2(\vec{x} - \vec{x}_*)\, dx^1 \wedge dx^2\right)\right>.
\end{align}
Mass $m \neq 0$ and spin $s$ of a particle are encoded in $c_0 = m\, J_0 + s\, P_0 \in \mathfrak{g}$, while a gauge group element $h$ acting via $h c_0 h^{-1} = {\bf p} + {\bf j}$ determines its momentum ${\bf p} = p^\mu J_\mu$ and (generalized) angular momentum ${\bf j} = j^\mu P_\mu$. 

Furthermore, $A_t$ acts as a Lagrange multiplier imposing a constraint on the curvature of spatial connection ($F_S = dA_S + [A_S,A_S]$):
\begin{align}\label{eq:10.07}
F_{\cal S} = 8\pi G\, h c_0 h^{-1} \delta^2(\vec{x} - \vec{x}_*)\, dx^1 \wedge dx^2\,.
\end{align}
The definition (\ref{eq:10.02}) leads to the relation $F_{\cal S} = R_{\cal S} + T_{\cal S} + C_{\cal S}$, from which it follows that the spatial Riemann curvature and torsion are given by
\begin{align}\label{eq:10.08}
R_{\cal S} &= -C_{\cal S} + 8\pi G\, {\bf p}\, \delta^2(\vec{x} - \vec{x}_*)\, dx^1 \wedge dx^2\,, \nonumber\\ 
T_{\cal S} &= 8\pi G\, {\bf j}\, \delta^2(\vec{x} - \vec{x}_*)\, dx^1 \wedge dx^2\,,
\end{align}
i.e.\! they vanish (on the constant background $R_{\cal S} = -C_{\cal S}$, where $C_{\cal S}$ is a $\Lambda$-dependent term) everywhere except singularities at the particle's worldline.

\subsection{Structure of the gauge algebras and groups}
For any value of $\Lambda$, the brackets of generators of $\mathfrak{g}$ have the form
\begin{align}\label{eq:10.10}
[J_\mu,J_\nu] = \epsilon_{\mu\nu}^{\quad\!\sigma} J_\sigma\,, \quad [J_\mu,P_\nu] = \epsilon_{\mu\nu}^{\quad\!\sigma} P_\sigma\,, \quad [P_\mu,P_\nu] = -\Lambda\, \epsilon_{\mu\nu}^{\quad\!\sigma} J_\sigma\,.
\end{align}
The identification $P_\mu \equiv \theta J_\mu$, $\theta^2 = -\Lambda$ allows \cite{Meusburger:2008qr} to represent each $\mathfrak{g}$ as an extension of $\mathfrak{o}(2,1)$ over the ring of elements $a + \theta b$, $a,b \in \mathbbm{R}$. 

On the other hand, introducing a vector ${\bf n} \in \mathbbm{R}^{2,1}$ (timelike for $\Lambda > 0$, lightlike for $\Lambda = 0$ or spacelike for $\Lambda < 0$), ${\bf n}^2 = \Lambda$, one may define
\begin{align}\label{eq:10.11}
S_\mu := P_\mu + \epsilon_{\mu\nu}^{\quad\!\sigma} n^\nu J_\sigma\,,
\end{align}
which are generators of the so-called $\mathfrak{an}_{\bf n}(2)$ algebra. 
Then, we obtain
\begin{align}\label{eq:10.12}
[J_\mu,J_\nu] &= \epsilon_{\mu\nu}^{\quad\!\sigma} J_\sigma\,, \qquad 
[J_\mu,S_\nu] = \epsilon_{\mu\nu}^{\quad\!\sigma} S_\sigma + 
n_\nu J_\mu - \eta_{\mu\nu} n^\sigma J_\sigma\,, \nonumber\\ 
[S_\mu,S_\nu] &= n_\mu S_\nu - n_\nu S_\mu\,,
\end{align}
and the Lie group corresponding to $\mathfrak{g}$ will reveal an interesting structure. 

Namely, for each gauge group $G$, $g \in G$ has the Iwasawa decomposition:
\begin{align}\label{eq:10.13}
g &= \mathfrak{u}\, \mathfrak{s}\ \in\ {\rm SL}(2,\mathbbm{R}) \vartriangleright\!\!\vartriangleleft {\rm AN}_{\bf n}(2)\,, 
\end{align}
under the condition $s_3 + \tfrac{1}{2} {\bf n} \cdot {\bf s} > 0$ (in the parametrization (\ref{eq:10.15})), and/or
\begin{align}\label{eq:10.14}
g = \mathfrak{r}\, \mathfrak{v}\ \in\ {\rm AN}_{\bf n}(2) \vartriangleright\!\!\vartriangleleft {\rm SL}(2,\mathbbm{R})\,, 
\end{align}
under the condition $r_3 - \tfrac{1}{2} {\bf n} \cdot {\bf r} > 0$ \cite{Meusburger:2008qr}. Instead of 3D Lorentz group ${\rm SO}^\uparrow(2,1)$ (generated by $J_\mu$), we use here its double cover, ${\rm SL}(2,\mathbbm{R})$ or ${\rm SU}(1,1)$. The double product $\vartriangleright\!\!\vartriangleleft$ means that both components act on each other in a complicated manner. In the case of Poincar\'{e} group and ${\bf n} = {\bf 0}$, ${\rm AN}_{\bf n}(2)$ reduces to $\mathbbm{R}^{2,1}$ and $\vartriangleright\!\!\vartriangleleft$ to $\vartriangleright\!\!<$ (i.e. the semidirect product). 

The conditions below eqs. (\ref{eq:10.13}), (\ref{eq:10.14}) are given in terms of the ``quaternionic'' parametrization ($\mathbbm{1}$ and $\{J_\mu\}$ form a basis of pseudo-quaternions):
\begin{align}\label{eq:10.15}
\mathfrak{u} &= u_3 \mathbbm{1} + u^\mu J_\mu\,, \quad u_3^2 = 1 - \tfrac{1}{4} {\bf u}^2\,; \nonumber\\
\mathfrak{s} &= s_3 \mathbbm{1} + s^\mu S_\mu\,, \quad s_3^2 = 1 + \tfrac{1}{4} ({\bf n} \cdot {\bf s})^2\,,
\end{align}
where $u_3, u^\mu, s_3, s^\mu \in \mathbbm{R}$ (and analogously for $\mathfrak{v}$ and $\mathfrak{r}$). Explicitly, ${\rm SL}(2,\mathbbm{R})$ can be expressed in a $2 \times 2$ representation of its algebra:
\begin{align}\label{eq:10.16}
J_0 = \frac{1}{2} \left(
\begin{array}{cc}
0 & 1 \\ 
-1 & 0
\end{array}
\right), \quad 
J_1 = \frac{1}{2} \left(
\begin{array}{cc}
1 & 0 \\ 
0 & -1
\end{array}
\right), \quad 
J_2 = \frac{1}{2} \left(
\begin{array}{cc}
0 & 1 \\ 
1 & 0
\end{array}
\right)
\end{align}
while the representation of ${\rm AN}_{\bf n}(2)$ is obtained by applying (\ref{eq:10.16}) to the formula $S_\mu = \theta J_\mu + \epsilon_{\mu\nu}^{\quad\!\sigma} n^\nu J_\sigma$. In our context, this approach is more useful than to consider an exponential map $g = \exp(\xi^\mu J_\mu) \exp(\varepsilon^\mu S_\mu)$ to define a $4 \times 4$ representation of ${\rm SO}^\uparrow(2,1) \vartriangleright\!\!\vartriangleleft {\rm AN}_{\bf n}(2)$ via (e.g., for $\Lambda > 0$):
\begin{align}\label{eq:10.17}
J_0 &= \left(
\begin{array}{cccc}
0 & 0 & 0 & 0 \\ 
0 & 0 & 1 & 0 \\ 
0 & -1 & 0 & 0 \\ 
0 & 0 & 0 & 0
\end{array}
\right), & 
S_0 &= \sqrt{\Lambda} \left(
\begin{array}{cccc}
0 & 0 & 0 & -1 \\ 
0 & 0 & 0 & 0 \\ 
0 & 0 & 0 & 0 \\ 
-1 & 0 & 0 & 0
\end{array}
\right), \nonumber\\ 
J_1 &= \left(
\begin{array}{cccc}
0 & 0 & -1 & 0 \\ 
0 & 0 & 0 & 0 \\ 
-1 & 0 & 0 & 0 \\ 
0 & 0 & 0 & 0
\end{array}
\right), & 
S_1 &= \sqrt{\Lambda} \left(
\begin{array}{cccc}
0 & 1 & 0 & 0 \\ 
1 & 0 & 0 & 1 \\ 
0 & 0 & 0 & 0 \\ 
0 & -1 & 0 & 0
\end{array}
\right), \nonumber\\ 
J_2 &= \left(
\begin{array}{cccc}
0 & 1 & 0 & 0 \\ 
1 & 0 & 0 & 0 \\ 
0 & 0 & 0 & 0 \\ 
0 & 0 & 0 & 0
\end{array}
\right), & 
S_2 &= \sqrt{\Lambda} \left(
\begin{array}{cccc}
0 & 0 & 1 & 0 \\ 
0 & 0 & 0 & 0 \\ 
1 & 0 & 0 & 1 \\ 
0 & 0 & -1 & 0
\end{array}
\right).
\end{align}
The representation (\ref{eq:10.17}) also does not preserve the relation $P_\mu \equiv \theta J_\mu$.

\section{Effective particle actions and properties of particles}
The Alekseev-Malkin construction is a way \cite{Meusburger:2005by} to integrate the gravitational degrees of freedom. To this end, we decompose space ${\cal S}$ into a disc containing the particle ${\cal D}$ and the empty region ${\cal E}$, sharing the boundary $\Gamma$. From the constraint (\ref{eq:10.07}), it follows that the connection on ${\cal E}$ has the form
\begin{align}\label{eq:20.01}
A_{\cal S}\big|_{\cal E} = \gamma d\gamma^{-1}\,,
\end{align}
while on ${\cal D}$ (in coordinates $\rho \in (0,1]$, $\phi \in [0,2\pi)$) it is
\begin{align}\label{eq:20.02}
A_{\cal S}\big|_{\cal D} = \bar\gamma\, 4G c_0 d\phi\, \bar\gamma^{-1} + 
\bar\gamma d\bar\gamma^{-1}\,, \quad \bar\gamma(\rho\! =\! 0) = h\,.
\end{align}
The continuity of $A_{\cal S}$ across $\Gamma$ leads to the sewing condition (in particular, $\gamma \in G$ has a jump at $\phi = 0 \cong 2\pi$, due to the conical defect on ${\cal D}$)
\begin{align}\label{eq:20.03}
\gamma^{-1}\big|_\Gamma = \alpha\, e^{4G c_0 \phi} 
\bar\gamma^{-1}\big|_\Gamma\,, \quad d\alpha = 0\,.
\end{align}
Applying Iwasawa decompositions (\ref{eq:10.13}) to eq.~(\ref{eq:20.03}) and performing further manipulations, we can express our Lagrangian as a boundary integral
\begin{align}\label{eq:20.04}
L &= \kappa \int_\Gamma \left<\partial_0\left(\bar{\mathfrak{u}}^{-1} 
\mathfrak{u}\right) \mathfrak{u}^{-1} \bar{\mathfrak{u}} 
\left(d\bar{\mathfrak{s}}\, \bar{\mathfrak{s}}^{-1} - 
\bar{\mathfrak{s}}\, \frac{c_0}{2\pi\kappa} d\phi\, \bar{\mathfrak{s}}^{-1}\right) + \frac{c_0}{2\pi\kappa} d\phi\, \bar{\mathfrak{s}}^{-1} 
\dot{\bar{\mathfrak{s}}}\right>,
\end{align}
where $\kappa \equiv \frac{1}{8\pi G}$. However, apart from the case of $\Lambda = 0$ and ${\bf n} = {\bf 0}$, the formula for $\mathfrak{u}$ is too unwieldy to proceed further \cite{Trzesniewski:2018cy}.

\subsection{Results for Poincar\'{e} group}

For this particular group, we are able \cite{Arzano:2014by} to perform the integration in eq.~(\ref{eq:20.04}) and (after fixing the gauge as $\gamma(0) = 1$) obtain the effective Lagrangian, which agrees with the one derived in the metric formalism \cite{Matschull:1997qy}:
\begin{align}\label{eq:20.05}
L &= \kappa\, \big(\dot\Pi^{-1} \Pi\big)_{\!\mu} x^\mu + 
s\, \tfrac{1}{2} \epsilon_{0\mu}^{\quad\!\nu} \dot \Lambda^\mu_{\ \sigma}(\bar{\mathfrak{u}}^{-1}) \Lambda^\sigma_{\ \nu}(\bar{\mathfrak{u}})\,, \nonumber\\
&= -\tfrac{1}{2} \epsilon_\nu^{\ \sigma\mu} \dot \Lambda^\nu_{\ \varrho}(\bar{\mathfrak{u}}) \Lambda^\varrho_{\ \sigma}(\bar{\mathfrak{u}}^{-1})\, \Upsilon_\mu\,.
\end{align}
It is given in terms of the particle's momentum $\Pi \equiv \bar{\mathfrak{u}}\, e^{\frac{m}{\kappa} J_0} \bar{\mathfrak{u}}^{-1} \in {\rm SL}(2,{\mathbbm R})$ and position ${\bf x} \equiv \bar{\mathfrak{u}}\, \bar{\bf s}\, \bar{\mathfrak{u}}^{-1} \in \mathbbm{R}^{2,1}$, or angular momentum $\Upsilon \equiv \kappa\, ({\bf x} - \Pi\, {\bf x}\, \Pi^{-1} + \bar{\mathfrak{u}}\, \tfrac{s}{\kappa} P_0\, \bar{\mathfrak{u}}^{-1}) \in {\mathbbm R}^{2,1}$, as well as a Lorentz transformation $\Lambda^\mu_{\ \nu}(\bar{\mathfrak{u}})$ corresponding to $\bar{\mathfrak{u}}$, $\Lambda^\mu_{\ \nu}(\bar{\mathfrak{u}})\, J_\mu := \bar{\mathfrak{u}}\, J_\nu \bar{\mathfrak{u}}^{-1}$. 

Parallel transport around the particle is described by holonomy of the connection $A_{\cal S}$ along the boundary $\Gamma$, which is a gauge group element
\begin{align}\label{eq:20.07}
{\cal P}\, e^{\int_\Gamma A_{\cal S}} &= \gamma(\phi\! =\! 0)\, \gamma^{-1}(\phi\! =\! 2\pi) = \Pi \left(\mathbbm{1} + \tfrac{1}{\kappa} \Pi^{-1} \Upsilon \Pi\right), 
\end{align}
It shows that the (extended) momentum space is indeed ${\rm SL}(2,{\mathbbm R})$, i.e. 3D anti-de Sitter manifold. Using the parametrization $\Pi = p_3 \mathbbm{1} + \frac{1}{\kappa}\, p^\mu J_\mu$ (cf. eq.~(\ref{eq:10.15})), we uncover deformations of the mass shell condition
\begin{align}\label{eq:20.08}
p_\mu p^\mu = 4\kappa^2 \sin^2\tfrac{m}{2\kappa}
\end{align}
and of the angular momentum
\begin{align}\label{eq:20.09}
\Upsilon^\mu = p_3\, \epsilon^\mu_{\ \nu\sigma} x^\nu p^\sigma + 
\tfrac{1}{2\kappa} \left(x^\mu p_\nu p^\nu - x^\nu p_\nu p^\mu\right) + \tfrac{s}{m}\, p^\mu\,.
\end{align}
However, variation of the action determined by the Lagrangian (\ref{eq:20.05}) still leads to the conservation law $\dot j_\mu = 0$, while imposing eq.~(\ref{eq:20.08}) as a constraint, we find that the equations of motion are also unaffected (cf. eq.~(\ref{eq:20.14}) below). 

Eqs.~(\ref{eq:20.08}) and (\ref{eq:20.09}) for a free particle are recovered in the limit of $\kappa \to \infty$. Similarly, the Lagrangian (\ref{eq:20.05}) can be reduced to the free-particle case
\begin{align}\label{eq:20.06}
L = p_\mu \dot x^\mu + 
s\, \tfrac{1}{2} \epsilon_{0\mu}^{\quad\!\nu} \dot \Lambda^\mu_{\ \sigma}(\bar{\mathfrak{u}}^{-1}) \Lambda^\sigma_{\ \nu}(\bar{\mathfrak{u}}) = \left<c_0 \alpha^{-1} \dot \alpha\right>,
\end{align}
coinciding with the second term in eq.~(\ref{eq:10.06}) (the spin term may seem unusual but has the same form as for a particle in 4D \cite{Sousa:1990os}).


\subsection{Generalization to multiple particles}\label{seq:3.2}
In a $n$-particle case (if $n \geq 3$, ${\cal S}$ can be a closed surface even for $\Lambda \leq 0$, while for $\Lambda > 0$, topology forces $n$ to be even), the Lagrangian (\ref{eq:10.06}) becomes
\begin{align}\label{eq:20.10}
L &= \frac{1}{16\pi G} \int_{\cal S} \left<\dot A_{\cal S} \wedge A_{\cal S}\right> - 
\sum_{i=1}^n \left<c_{(i)} h_i^{-1} \dot h_i\right> \nonumber\\ 
&+ \int_{\cal S} \left<A_0 \left(\frac{1}{8\pi G} F_{\cal S} - 
\sum_{i=1}^n h_i c_{(i)} h_i^{-1} \delta^2(\vec{x} - \vec{x}_i)\, dx^1 \wedge dx^2\right)\right>,
\end{align}
where $c_{(i)} = m_{(i)} J_0 + s_{(i)} P_0$. Dividing space ${\cal S}$ into $n$ particle discs ${\cal D}_i$ and an empty polygon ${\cal E}$, with the common boundary $\Gamma = \bigcup_i \Gamma_i$, we can follow the earlier example of a single particle and, if $\Lambda = 0$, derive the effective Lagrangian for each $i$. Then, we impose the continuity of $\gamma$ at the vertices of ${\cal E}$ (except a jump between $\Gamma_n$ and $\Gamma_1$) and finally obtain \cite{Trzesniewski:2018cy}
\begin{align}\label{eq:20.11}
L 
&= \sum_{i=1}^n \left(\kappa\, \big(\dot\Pi_i^{-1} \Pi_i\big)_{\!\mu} \left({\bf x}_i\right)^\mu + s_{(i)} \tfrac{1}{2} \epsilon_{0\mu}^{\quad\!\nu} \dot \Lambda^\mu_{\ \sigma}(\bar{\mathfrak{u}}_i^{-1}) \Lambda^\sigma_{\ \nu}(\bar{\mathfrak{u}}_i)\right. \nonumber\\
&\left.- \big(\partial_0(\Pi_{i-1}^{-1} \ldots \Pi_1^{-1})\, \Pi_1 \ldots \Pi_{i-1} \big)_{\!\mu} \left(\Upsilon_i\right)^\mu\right).
\end{align}
Each $i$'th summand differs from the case (\ref{eq:20.05}) by an extra ``interaction term'' in the second line. However, after adding the mass-shell conditions (analogous to eq.~(\ref{eq:20.08})), we are still led to the free-particle equations of motion
\begin{align}\label{eq:20.14}
\dot x_{(i)}^\mu = \lambda_{(i)} \cos\frac{m_{(i)}}{2\kappa}\, p_{(i)}^\mu\,, \qquad 
\dot p_{(i)}^\mu = 0\,,
\end{align}
where $\lambda_{(i)} \cos\frac{m_{(i)}}{2\kappa}$ is just a rescaled Lagrange multiplier. The lack of actual interactions reflects the topological nature of the theory. (Interestingly, if ${\cal S}$ is an open surface, the spatial infinity can be contracted into a fictitious extra particle, capturing the boundary conditions \cite{Meusburger:2005by}.) 

A holonomy calculated by circumventing $j \leq n$ particles is given by
\begin{align}\label{eq:20.12}
{\cal P}\, e^{\int_{\Gamma(j)} A_{\cal S}} &= \gamma(\phi_1\! =\! 0)\, \gamma^{-1}(\phi_j\! =\! 2\pi) = \Pi_1 \ldots \Pi_j \nonumber\\ 
&\cdot \left(\mathbbm{1} + \tfrac{1}{\kappa} \Pi_j^{-1} \ldots \Pi_1^{-1} \Upsilon_1 \Pi_1 \ldots \Pi_j + \ldots + \tfrac{1}{\kappa} \Pi_j^{-1} \Upsilon_j \Pi_j\right).
\end{align}
In particular, $\Pi_1 \ldots \Pi_j$ is the total momentum of $j$ particles. Holonomies are not invariant under permutations of particles $(g_i,g_{i+1}) \rightarrow (g_{i+1},g_i)$ but under their right- or left-handed braids (here $g_i \equiv \Pi_i (\mathbbm{1} + \Upsilon_i)$)
\begin{align}\label{eq:20.13}
(g_i,g_{i+1}) &\rightarrow (g_{i+1},g_{i+1}^{-1} g_i g_{i+1})\,, \nonumber\\ 
(g_i,g_{i+1}) &\rightarrow (g_i g_{i+1} g_i^{-1},g_i)\,,
\end{align}
which is a straightforward consequence of living in 2 spatial dimensions and should heavily influence the predictions of the quantum theory \cite{Arzano:2014by}.

\subsection{Contraction to a deformed Carrollian theory} 
Poincar\'{e} (gauge) group is equivalent to the limit of a group contraction ${\rm SL}(2,\mathbbm{R}) \vartriangleright\!\!\vartriangleleft {\rm AN}_{\bf n}(2) \to {\rm SL}(2,\mathbbm{R}) \vartriangleright\!\!< \mathbbm{R}^{2,1}$. On the other hand, one may try to consider Chern-Simons theory for the complementary contraction ${\rm SL}(2,\mathbbm{R}) \vartriangleright\!\!\vartriangleleft {\rm AN}_{\bf n}(2) \to \mathbbm{R}^{2,1} >\!\!\vartriangleleft {\rm AN}_{\bf n}(2)$ (with the help of appropriate rescalings by $\Lambda$). We discovered \cite{Kowalski:2014dy,Trzesniewski:2018cy} that it leads -- but only for $\Lambda > 0$ -- from eq. (\ref{eq:10.06}) to the effective particle Lagrangian
\begin{align}\label{eq:20.15}
L = \kappa\, \big(\Pi \dot\Pi^{-1}\big)_{\!\mu} x^\mu + 
s\, \big(\bar{\mathfrak{s}}^{-1} \dot{\bar{\mathfrak{s}}}\big)_0\,,
\end{align}
with $\Pi \equiv \bar{\mathfrak{s}}\, e^{\frac{m}{\kappa} P_0} \bar{\mathfrak{s}}^{-1} \in {\rm AN}(2)$ and ${\bf x} \equiv \bar{\mathfrak{u}} \in \mathbbm{R}^{2,1}$. The derivation follows the same steps as for eq.~(\ref{eq:20.05}) but with ${\rm AN}_{\bf n}(2)$ playing the role of ${\rm SL}(2,\mathbbm{R})$, while mass and spin are swapped due to the rescalings. 

Calculation of the holonomy leads to the result analogous to eq.~(\ref{eq:20.07}) (but we do not define the quantity $\Upsilon$ here, since its meaning is unclear),
\begin{align}\label{eq:22.08}
{\cal P}\, e^{\int_\Gamma A_{\cal S}} = \gamma(\phi = 0)\, \gamma^{-1}(\phi = 2\pi) = \left(\mathbbm{1} + \left(1 - {\rm Ad}(\Pi)\right) {\bf x} + \tfrac{s}{\kappa} J_0\right) \Pi\,.
\end{align}
$\Pi$ is thus interpreted as momentum. Meanwhile, in terms of the parametrization $\Pi = e^{\frac{1}{\kappa} p^a S_a} e^{\frac{1}{\kappa} p^0 S_0}$, we find the mass shell condition $p_0 = m$ and, after including the latter as a constraint, rewrite the Lagrangian (\ref{eq:20.15}) as
\begin{align}\label{eq:22.13}
L = \dot x^0 p_0 + \dot x^a p_a + \kappa^{-1} x^a p_a \dot p_0 - \tfrac{\lambda}{2} \left(p_0^2 - m^2\right).
\end{align}
The spin term has now been omitted since it does not contribute to the equations of motion, which turn out to be given by
\begin{align}\label{eq:20.16}
\dot x^0 = \lambda\, m\,, \qquad \dot x^a = 0\,, \qquad \dot p^\mu = 0\,.
\end{align}
Therefore, surprisingly, the particle is always at rest. 

In the context of special relativity, a particle satisfies such equations in the limit of vanishing speed of light, known as the Carrollian limit \cite{Bacry:1968ps}. Eq.~(\ref{eq:22.13}) differs from the Lagrangian of a Carroll particle \cite{Bergshoeff:2014ds} by a term proportional to $\kappa^{-1}$, which means that, apparently, our model describes a particular deformed version of Carroll particle (see its symmetries in \cite{Kowalski:2014dy}). 

The generalization to multiple particles works as in the Poincar\'{e} case discussed in the previous subsection, see \cite{Trzesniewski:2018cy}.

\section{Towards deformed symmetries}
Quantisation of the theory of 3D gravity consists in the Hopf-algebraic deformation of the Poisson structure, which is determined by a given classical r-matrix associated with the gauge algebra. All such possible r-matrices have recently been classified \cite{Borowiec:2017bs}. One of the remaining open questions is whether the widely-studied $\kappa$-Poincar\'{e} algebra (associated with noncommutative $\kappa$-Minkowski space) actually plays a physical role here. This is the subject we studied in \cite{Kowalski:2020qs} but did not manage to discuss our results during the current conference due to the running out of time.

\end{document}